\documentclass[a4paper,english]{elsarticle}
\usepackage{graphicx}
\usepackage{rotating}
\usepackage{babel}
\begin{document}


\title {Background studies for the EDELWEISS dark matter experiment}

\author[]{The EDELWEISS Collaboration}

\author[irfu]{\\E.~Armengaud}
\author[ipnl]{C.~Augier}
\author[neel]{A.~Beno\^{\i}t}
\author[ipnl]{A.~Beno\^{\i}t}
\author[csnsm]{L.~Berg\'e}
\author[pek]{T.~Bergmann}
\author[iek,fzk]{J.~Bl$\mbox{\"u}$mer}
\author[csnsm]{A.~Broniatowski}
\author[jinr]{V.~Brudanin}
\author[ipnl]{B.~Censier}
\author[csnsm]{M.~Chapellier}
\author[ipnl]{F.~Charlieux}
\author[csnsm]{F.~Couedo}
\author[oxford]{P. Coulter}
\author[iek]{G.A.~Cox}
\author[ipnl]{M.~De~Jesus}
\author[csnsm,irfu]{J.~Domange}
\author[csnsm]{A.-A.~Drilien}
\author[csnsm]{L.~Dumoulin}
\author[fzk]{K.~Eitel}
\author[jinr]{D.~Filosofov}
\author[irfu]{N.~Fourches}
\author[ipnl]{J.~Gascon}
\author[irfu]{G.~Gerbier}
\author[irfu]{M.~Gros}
\author[oxford]{S. Henry}
\author[irfu]{S.~Herv\mbox{\'e}}
\author[iek]{G.~Heuermann}
\author[csnsm]{N.~Holtzer}
\author[ipnl]{A.~Juillard}
\author[pek]{M.~Kleifges}
\author[iek]{H.~Kluck}
\author[fzk]{V.~Kozlov}
\author[oxford]{H.~Kraus}
\author[sheffield]{V.A.~Kudryavtsev\corref{cor1}}
\author[csnsm]{H.~Le~Sueur}
\author[lsm]{P.~Loaiza\corref{cor1}}
\author[csnsm]{S.~Marnieros}
\author[pek]{A.~Menshikov}
\author[irfu]{X-F.~Navick}
\author[irfu]{C.~Nones}
\author[csnsm]{E.~Olivieri}
\author[iramis]{P.~Pari}
\author[irfu]{B.~Paul}
\author[csnsm]{O.~Rigaut}
\author[sheffield]{M.~Robinson}
\author[jinr]{S.~Rozov}
\author[ipnl]{V.~Sanglard}
\author[iek]{B.~Schmidt}
\author[ipnl]{S.~Scorza\fnref{pa-scorza}}
\fntext[pa-scorza]{Present address: 
Department of Physics, Southern Methodist University, Dallas, TX 75275, USA}
\author[fzk]{B.~Siebenborn}
\author[jinr]{S.~Semikh}
\author[pek]{D.~Tcherniakhovski}
\author[irfu]{A.S.~Torrento-Coello}
\author[ipnl]{L.~Vagneron}
\author[irfu,iek]{R.J.~Walker}
\author[pek]{M.~Weber}
\author[jinr]{E.~Yakushev}
\author[oxford]{X.~Zhang}

\address[irfu]{CEA, Centre d'Etudes Saclay, IRFU, 
91191 Gif-Sur-Yvette Cedex, France}
\address[ipnl]{IPNL, Universit\'{e} de Lyon, Universit\'{e} Lyon 1, 
CNRS/IN2P3, 4 rue E. Fermi 69622 Villeurbanne cedex, France}
\address[neel]{CNRS-N\'{e}el, 25 Avenue des Martyrs, 
38042 Grenoble cedex 9, France}
\address[csnsm]{CSNSM, Universit\'e Paris-Sud, IN2P3-CNRS, bat 108, 91405 Orsay,  France}
\address[pek]{Karlsruhe Institute of Technology, Institut f\"{u}r                    
Prozessdatenverarbeitung und Elektronik, 76021 Karlsruhe, Germany}
\address[iek]{Karlsruhe Institute of Technology,
Institut f\"{u}r Experimentelle Kernphysik, 76128 Karlsruhe, Germany}
\address[fzk]{Karlsruhe Institute of Technology,
Institut f\"ur Kernphysik, 76021 Karlsruhe, Germany}
\address[iramis]{CEA, Centre d'Etudes Saclay, 
IRAMIS, 91191 Gif-Sur-Yvette Cedex, France}
\address[jinr]{Laboratory of Nuclear Problems, JINR, Joliot-Curie 6, 
141980 Dubna, Moscow region, Russia}
\address[lsm]{Laboratoire Souterrain de Modane, CEA-CNRS, 
1125 route de Bardonn\`eche, 73500 Modane, France}
\address[oxford]{University of Oxford, Department of Physics, Keble Road, Oxford OX1 3RH, UK}
\address[sheffield]{Department of Physics and Astronomy, University of Sheffield, Hounsfield Road, Sheffield S3 7RH, UK}

\cortext[cor1]{Corresponding authors: v.kudryavtsev@sheffield.ac.uk, ploaiza@lsm.in2p3.fr}



\begin{abstract}
The EDELWEISS-II collaboration has completed a direct search for WIMP dark matter using cryogenic Ge detectors (400 g each) and 384 kg$\times$days of effective exposure. A cross-section of $4.4 \times 10^{-8}$ pb is excluded at 90\%~C.~L. for a WIMP mass of 85 GeV. The next phase, EDELWEISS-III, aims to probe spin-independent WIMP-nucleon cross-sections down to a few $\times10^{-9}$ pb. We present here the study of gamma and neutron background coming from radioactive decays in the set-up and shielding materials. We have carried out Monte Carlo simulations for the completed EDELWEISS-II setup with GEANT4 and normalised the expected background rates to the measured radioactivity levels (or their upper limits) of all materials and components. The expected gamma-ray event rate in EDELWEISS-II at 20-200 keV agrees with the observed rate of 82 events/kg/day within the uncertainties in the measured concentrations. The calculated neutron rate from radioactivity of 1.0-3.1 events (90\%~C.~L.) at 20-200 keV in the EDELWEISS-II data together with the expected upper limit on the misidentified gamma-ray events ($\le0.9$), surface betas ($\le0.3$), and muon-induced neutrons ($\le0.7$), do not contradict 5 observed events in nuclear recoil band. We have then extended the simulation framework to the EDELWEISS-III configuration with 800 g crystals, better material purity and additional neutron shielding inside the cryostat. The gamma-ray and neutron backgrounds in 24 kg fiducial mass of EDELWEISS-III have been calculated as 14-44 events/kg/day and 0.7-1.4 events per year, respectively. The results of the background studies performed in the present work have helped to select better purity components and improve shielding in EDELWEISS-III to further reduce the expected rate of background events in the next phase of the experiment.

\end{abstract}


\maketitle

\pagebreak

\section{Introduction}

The reduction and discrimination of the background is one of the most important tasks in any dark matter experiment as the signal rate expected from WIMPs is extremely low. EDELWEISS-II is a direct dark matter search experiment based on Ge bolometers. The combined measurement of the ionisation and heat in  a particle interaction allows the  rejection of the gamma background at the level of (3$\pm$1)$\times$10$^{-5}$ ~\cite{EDW2final}. Interleaved electrode design, recently developed by the collaboration ~\cite{interdigit1}, enables an efficient rejection (6$\times$10$^{-5}$ ~\cite{interdigit2}) of near-surface interactions. Using 10 detectors representing a total mass of  4 kg and with a total effective exposure of 384 kg$\times$days, EDELWEISS-II has recently published its final WIMP search result ~\cite{EDW2final}. A cross-section of 4.4 x 10$^{-8}$ pb has been excluded at 90\%~C.~L. for a WIMP mass of 85 GeV/$c^2$. 
To reach the sensitivity to WIMP-nucleon cross-section significantly below 10$^{-8}$ pb in the next phase of the experiment, the background has to be further reduced. 

The sources of background are  neutrons,  gamma-rays and surface beta contaminants. Neutrons may be induced by cosmic-ray muons or generated by the decay of the natural radioactive elements present in the cavern walls and in the set-up components. Details on the muon-induced neutron studies using the EDELWEISS-II setup are given in Ref.~\cite{klaus}, an additional liquid scintillator detector
dedicated to the measurement of muon-induced neutrons is described in Ref.~\cite{veto}. Gamma-rays and beta contaminants are produced by the radioactivity in the construction materials. Surface events induced by surface contaminants are discriminated using the interleaved electrodes. Ref. ~\cite{EDW2final} gives details on the surface event background.
We present in this paper studies of the gamma-ray and neutron background coming from radioactive decays in the set-up and shielding of EDELWEISS-II and EDELWEISS-III. Extensive Monte Carlo simulations have been performed and combined with radiopurity measurements of all materials. These background studies have been used for optimisation of the configuration of the next stage WIMP search experiment at Modane -- EDELWEISS-III. 

\section{Experimental set-up and simulations}
EDELWEISS-II is located in the Laboratoire Souterrain de Modane (LSM) where the rock overburden of 4800 m w.e. reduces the cosmic muon flux down to about 5 muons/m$^2$/day~\cite{klaus}. The environmental gamma-ray flux below 4 MeV is dominated by natural radioactivity in the rock and concrete. The uranium, thorium and potassium concentrations have been reported in \cite{chazal}: 0.84$\pm$ 0.2 ppm and 1.9 $\pm$ 0.2 ppm of $^{238}$U, 2.45$\pm$ 0.2 ppm and 1.4 $\pm$ 0.2 ppm of $^{232}$Th, 230$\pm$30 Bq/kg and 77.3$\pm$13 Bq/kg of K in the rock and concrete, respectively. The neutron flux above 1 MeV is about 10$^{-6}$ n/cm$^{2}$/s \cite{fiorucci07}.
The radon level in the laboratory is $\sim$20 Bq/m$^{3}$ thanks to a ventilation system renewing the entire laboratory volume 1.5 times per hour. Further reduction of the radon level (down to $\sim$20 mBq/m$^3$) inside the shielding is achieved by the radon trap facility.

\begin{figure}[htb]
\begin{minipage}{15pc}
\includegraphics[width=18pc]{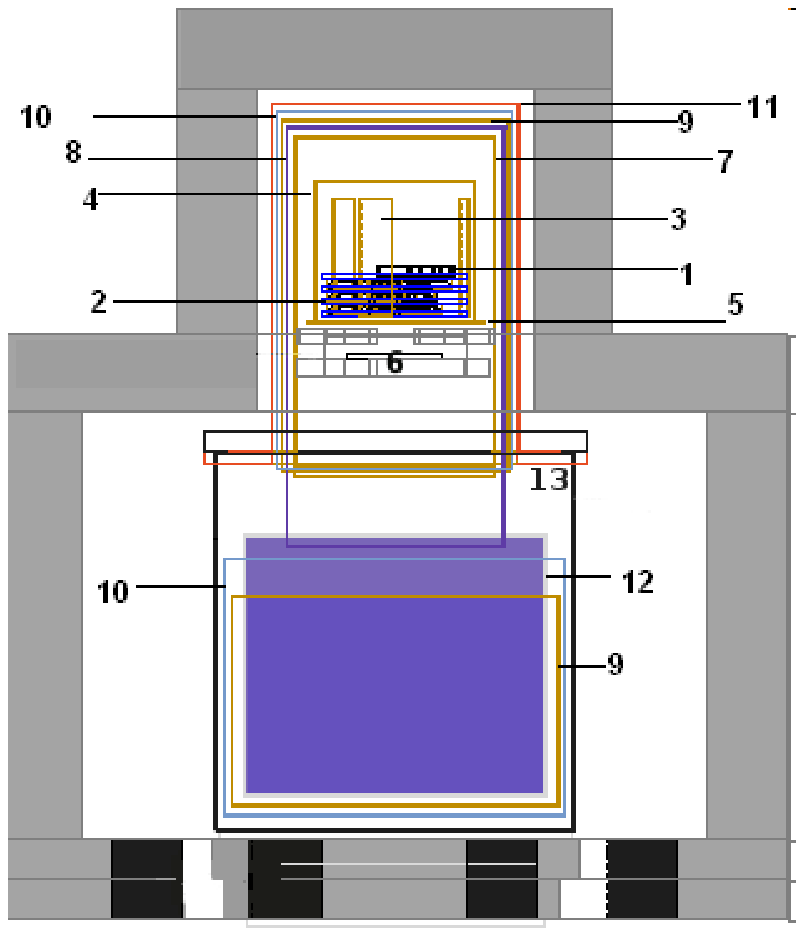}
\end{minipage}\hspace{4pc}
\begin{minipage}{15pc}
\includegraphics[width=18pc]{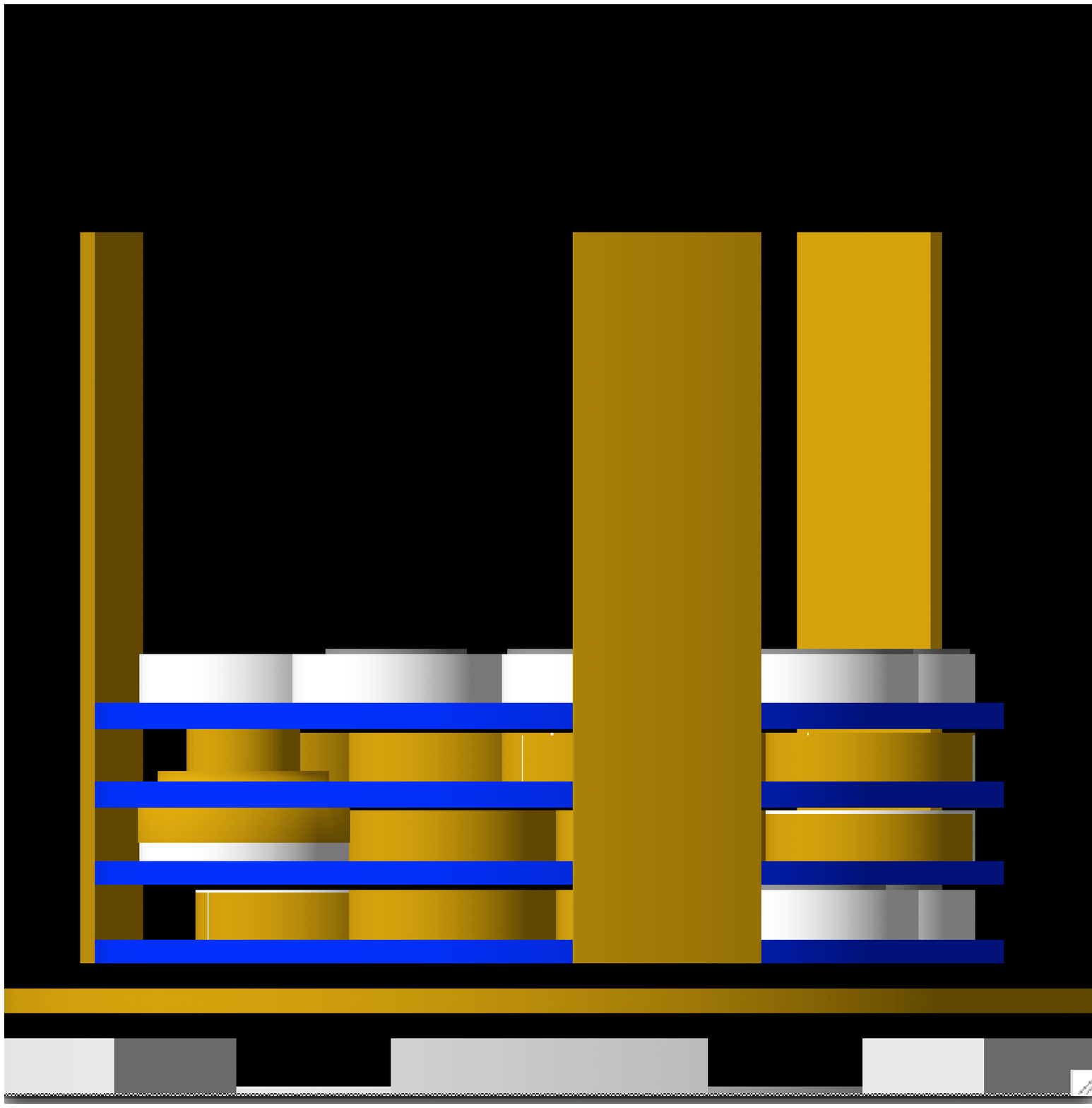}
\end{minipage} 
\caption{\label{fig:setup} (Color online) The GEANT4 geometry of the EDELWEISS-II set-up. Left : 1 -- germanium detectors with casings, 2 -- copper disks supporting Ge detectors (10 mK), 3 -- support bars for the copper disks (10 mK), 4 -- 10 mK thermal screen, 5 -- 10 mK thick plate supporting inner detector components, 6 -- internal roman lead shielding, 7 -- 1K thermal screen, 8 -- 4.2K thermal screen, 9 -- 40K thermal screen, 10 -- 100K thermal screen, 11 -- 300K vacuum chamber, 12 -- stainless steel liquid He reservoir, 13 -- stainless steel can. The outer lead shielding including modern and roman lead, is shown in grey. The outer polyethylene shielding and the muon veto are not shown. Right: zoom of the central part showing the germanium detectors with casings (dark yellow and grey) stacked on the copper disks (blue), the vertical support bars (dark yellow), the 10 mK thick plate (dark yellow) and, at the bottom, part of the internal roman lead shielding (grey).}
\end{figure}

EDELWEISS-II uses cryogenic germanium detectors installed in the 10 mK chamber of a dilution refrigerator specially designed for the experiment. Each detector is enclosed in an individual casing made of electrolytic copper of type CuC2 as termed by the manufacturer and characterised by high purity (99.99\% pure) and concentration of oxygen limited to 5 ppm. The radiopurity of this copper has been measured at LNGS (Italy) using gamma-spectrometry \cite{GeMPI} and the results are shown in Table 1. Only teflon (PTFE) is used to hold the detectors inside the casings in a design specially developed to obtain the lowest possible radioactive background \cite{XFN}.  
The detectors are arranged on disks supported by three vertical bars. The disks and the vertical bars are themselves supported by a thick plate at 10 mK and surrounded by a 10 mK thermal screen. The 10 mK plate also plays the role of shielding the Ge crystals from the radioactivity beneath the plate. The 10 mK plate and the 10 mK thermal screen will be referred to hereafter as the 10 mK chamber. The disks, the bars and the 10 mK chamber are made of electrolytic copper of type CuC1 (with oxygen concentration less than 1 ppm and purity of 99.95\%). The radiopurity of CuC1 copper has been measured at LSM. The results of the measurements are shown in Table~\ref{tab:radioactivities}.
To simulate the response of the detectors to various types of particles, the complete set-up has been implemented in the GEANT4 package \cite{Geant4}  as shown in Figure~\ref{fig:setup}.

Below the 10 mK plate, at 1K,  14 cm of roman lead shields the detectors from the gamma-rays induced by the radioactivity in the cold electronics, the dilution unit and other cryogenic parts. The dilution unit components are made of copper, stainless steel and silver. Four thermal screens at 1K, 4.2K, 40K, 100K and the vacuum chamber at 300K, all made of copper which has not been specially selected for its ultra-low radioactivity, complete the cryostat. Hereafter the thermal screens from 1K to 100K and the vacuum chamber at 300K will be referred to as `screens 7 to 11' according to the numbering in Figure 1. EDELWEISS-II uses coaxial cables from the detectors to room temperature. Resistors together with electrical connectors are installed at the 1K stage below the lead shielding. Cold JFETs are positioned at the 100K stage. The electronics to bias the JFETs, the DACs to bias the detectors, the final amplification, the anti-aliasing filter and the digitisation are all integrated in a single room-temperature module, called bolometer box, which is attached to the stainless steel can (see Figure 1 for details of the set-up, bolometer boxes are not shown). 

An 18 cm thick outer layer of modern lead shields the cryostat against ambient gamma-ray background. A 2 cm thick inner roman lead layer has been cast directly on the modern lead. An outer 50 cm thick polyethylene shielding protects the detector against ambient neutrons. The lead and polyethylene shielding is mounted on a mild steel structure with rails allowing the opening of the two halves of the shielding structure.
In addition an 100 m$^2$ plastic scintillator active muon veto surrounds the polyethylene \cite{veto}.

All materials used in the construction have been measured to assess their radioactive contaminations. Table~\ref{tab:radioactivities} shows
 a selection of the results. The CuC2 copper of the detector casings was purchased in 2006 and stored in LSM since then.
 A few samples of this copper have been exposed to cosmic rays for a few days during their transportation from LSM to LNGS for accurate measurements of radiopurity. 
Decay rates of cosmogenic isotopes (see Table ~\ref{tab:radioactivities}) agree with the assumption of a few day activation.

\begin{center}
\begin{table}[htb]
\caption{\label{tab:radioactivities}Radioactive contaminations in materials of the EDELWEISS-II set-up and shielding. All contaminations have been assessed by gamma-ray spectrometry, except for $^{238}$U and $^{232}$Th in lead and mild steel which have been measured by mass spectrometry, and $^{238}$U and $^{232}$Th in polyethylene measured by neutron activation. The radioactivity quoted for the dilution unit is based on measurement of individual components. }

\centering
{\footnotesize
\begin{tabular}{@{}*{8}{l}}
\hline
\\
Component/                     & Mass      & \multicolumn{5}{l}{Radioactivity in materials (mBq/kg or mBq/unit$^{\star}$) } \\
\\
Material                     & (kg)        & $^{226}$Ra    &  $^{228}$Th  &  $^{60}$Co& $^{40}$K              &  Other radionuclides \\
\\
\hline
\\
Detector holders/PTFE         &  0.02           & $<$7     & $<$5          & $<$20     & $<$100                & $^{210}$Pb$<$80 \\
\\
Electrodes/Al                 & $<$3$\cdot$10$^{-5}$ &  0.27$\pm$0.19       &  1.4$\pm$0.2        &  -        & 1.1 $\pm^{0.2}_{0.1}$  &$^{26}$Al: 0.38$\pm^{0.19}_{0.14}$ \\
\\
Detector casings/     &   3     & 0.025        & 0.033       & 0.038       & $<$0.39  & $^{238}$U$<$ 1.4, $^{235}$U$<$ 0.9  \\
CuC2 copper$^{a}$            &                 &  $\pm$0.015   & $\pm$0.016  & $\pm$0.010  &         & $^{54}$Mn: 0.024$\pm$0.010$^{b}$  \\
\\
Disks, bars, 10 mK chamber/                   &  90              &  $<$1           &  $<$0.7        &  $<$1        &     $<$110    & $^{210}$Pb:180$\pm$140  \\
CuC1 copper             &            &                &               &              &                &      \\       
\\ 
Screens 7 to 11/copper            &  320       & $<$3           & $<$2           & $<$2           &  $<$25        &        \\
\\
Dilution unit$^{\star}$               &   $\approx$1  &  $<$20       & $<$20           &  $<$20       & $<$100         & $^{108}$Ag:331 $\pm$32     \\
\\
1K connectors    & 0.32         &  644$\pm$65           &1353$\pm$138            & $<$25        & 1181 $\pm$197  &$^{238}$U:1994$\pm$204       \\
\\
Coaxial cables          & 1.4          & 10$\pm$7       & $<$6                 & $<$8        & 120 $\pm$60              & $^{210}$Pb$<$110   \\
\\
Bolometer boxes$^{\star}$           & 50 units       & 331$\pm$17  & 235$\pm$13 &       -         &    340$\pm$40  &  $^{238}$U:134 $\pm^{65}_{15}$   \\
(warm electronics)        &                 &   &     &                &         &   $^{210}$Pb :1019$\pm$56  \\ 
\\             
\hline
\\
Roman lead shield        & $\approx$120       &  $<$0.3          &   $<$0.3           &   -              & $<$1.3        & $^{210}$Pb$<$120   \\
\\
Modern lead shield       & 30000             &  $<$3           &  $<$1               &     -            &   -             & $^{210}$Pb: (24$\pm$1)$\times$10$^{3}$    \\
                         &                    &                 &                    &                  &                & $^{238}$U$<$ 0.01 ppb   \\
\\
Polyethylene shield      & 40000             & 5$\pm$1              &  $<$2            &   $<$3         & 16$\pm$2       & $^{238}$U:1 ppb,$^{232}$Th:0.1 ppb     \\
\\
Mild steel support       & 8600              &    -               &     -             &    -             &   -             & $^{238}$U$<$ 0.01 ppb \\
                         &                   &                   &                 &                  &                & $^{232}$Th$<$ 0.01 ppb     \\
\\
\hline
\end{tabular}}
\\

$^{a}$ CuC2 copper has been measured at LNGS with the GeMPI detector \cite{GeMPI}. 
$^{b}$ The activities of short-lived cosmogenic isotopes in CuC2 copper correspond to (10$\pm$2) days of exposure. 
\end{table}
\end{center}

\section{Gamma background}

The Monte Carlo simulation was based on the GEANT4 code with the Low Energy Electromagnetic Interactions physics list. Cross-sections are determined from evaluated data (EPDL97, EEDL and EADL, stopping power data, binding energy based on data of Scofield) \cite{Geant4_manual}. The particle generator uses the GEANT Radioactivity Decay generator (GRDM), which was designed to handle all kinds of decays ($\alpha$, $\beta^{-}$, $\beta^{+}$, EC), the emission of the associated particles and energy distribution, the following de-excitation of the nucleus ($\gamma$, internal conversion) and the accompanying X-rays and Auger electrons ~\cite{Geant4_manual}. The GRDM generator takes into account the total energy loss occurring due to the cascade gamma emission. All emitted particles were followed in GEANT4 and energy depositions in the crystals were stored. Energy depositions occurring in the same crystal within the time window of 50 ms were summed together giving a single event. In a subsequent analysis the fiducial events have been defined in the same way as in real data and the fiducial volume cut was applied to the simulated events.

The decays of $^{226}$Ra, $^{228}$Ra, $^{60}$Co, $^{40}$K, $^{54}$Mn and $^{210}$Pb were simulated in the detector casings, the disks supporting the Ge detectors, the  bars supporting the disks,  the 10 mK chamber, the cryostat screens 7 to 11 , the dilution unit (as a block for simplicity), 1K connectors, the coaxial cables and the lead shielding (see Figure~\ref{fig:setup}). To simplify the simulation task $^{228}$Ra was assumed to be in equilibrium with $^{228}$Th. In addition, the cosmogenically induced isotopes  $^{68}$Ge and $^{65}$Zn in germanium crystals were considered and their activities were chosen to match the measured intensities of the lines at around 10 keV.

PTFE crystal holders, aluminium electrodes and other small parts located close to the crystals have too small mass to give a measurable contribution to the gamma-ray background but their contribution to the neutron background may be enhanced due to high ($\alpha$,n) cross-sections, in particular on aluminium and fluorine (see Section 4). The coaxial cables, the dilution unit and 1K connectors (Table~\ref{tab:radioactivities}) are located below the 14 cm thick lead plate and their contribution was found to be negligible compared to that of the cryostat screens 7 to 11 despite higher radioactivity levels. The gamma-ray background from bolometer boxes (warm electronics) was not simulated as they were located behind lead.
The contribution from $^{210}$Bi in the modern lead shielding is negligible but the energetic gammas of about 2.6 MeV from $^{228}$Th decay chain may reach the detectors, as shown in Table~\ref{tab:eventrate_le}. The background from rock and concrete was shown to be suppressed by several orders of magnitude due to the lead shielding around the cryostat. 

 As only upper limits were obtained in the radioactivity measurements for CuC1 copper and the copper of the screens 7 to 11, a $\chi^2$ minimisation with 10 free parameters was used to determine those contaminations. The 10 free parameters were: $^{226}$Ra and $^{228}$Ra in CuC1 copper (disks, bars and 10 mK chamber), $^{226}$Ra and $^{228}$Ra in copper of the screens 7 to 11 and $^{226}$Ra and $^{228}$Ra contamination at 300K (not shown in Table~\ref{tab:radioactivities}) which could be due to unaccounted radioactivity in cryogenic pipes, electronics, radon or uncontrolled impurities on the 300K vacuum chamber (6 parameters in total); cosmogenic $^{60}$Co and $^{54}$Mn in CuC1 and copper of the screens 7 to 11 (assumed to be the same), $^{40}$K in CuC1 and copper of the screens 7 to 11 (assumed to be the same) and $^{210}$Pb on the surface of the detector casings.
The radiopurity measurements reported in Table~\ref{tab:radioactivities} for CuC2 copper were used to calculate the gamma contributions from CuC2 copper parts. The upper limits for other copper parts were taken as upper bounds for the fitting procedure.

\begin{figure}[htb]
\begin{minipage}{18pc}
\includegraphics[width=18pc]{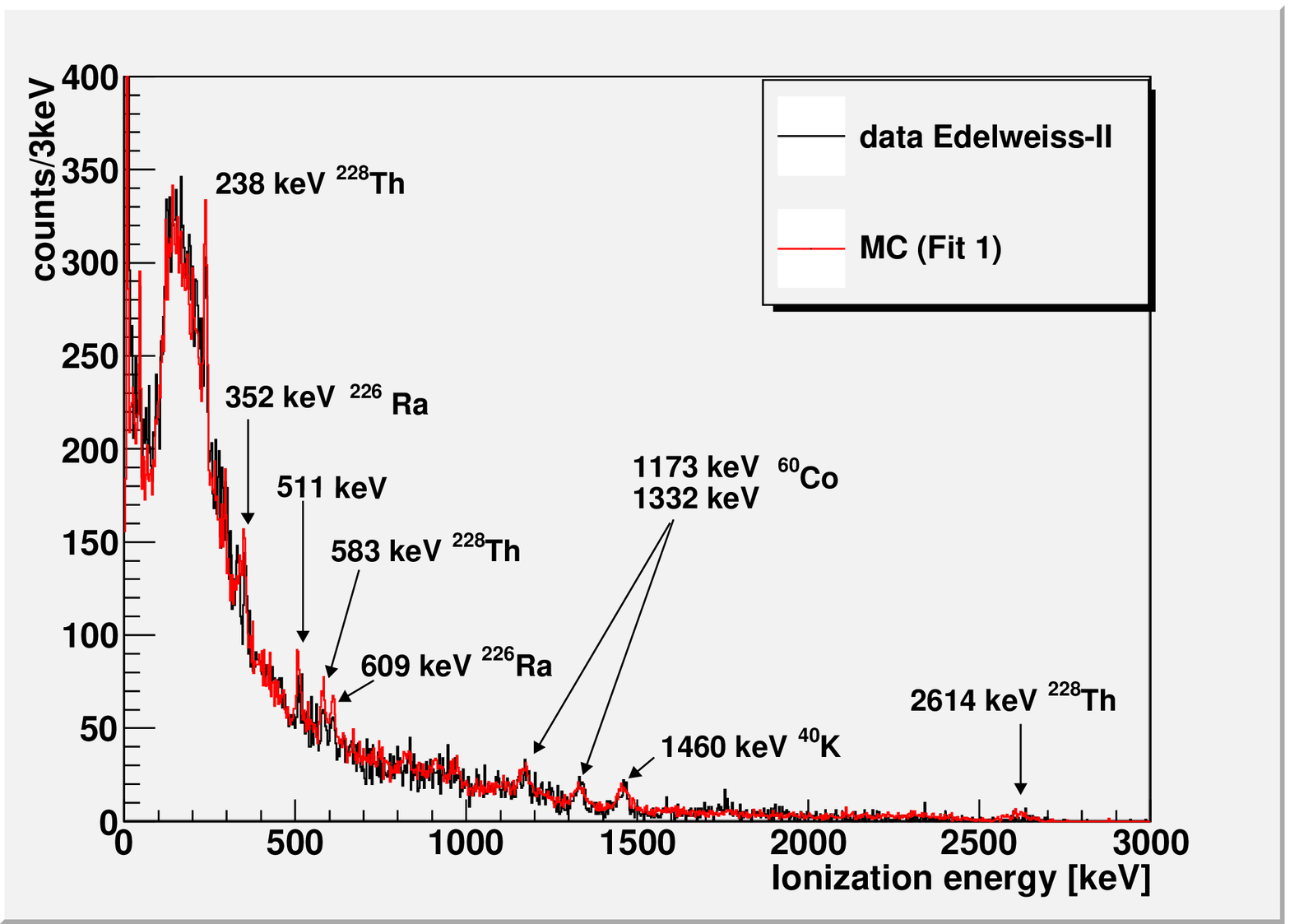}
\end{minipage}\hspace{1pc}
\begin{minipage}{18pc}
\includegraphics[width=18pc]{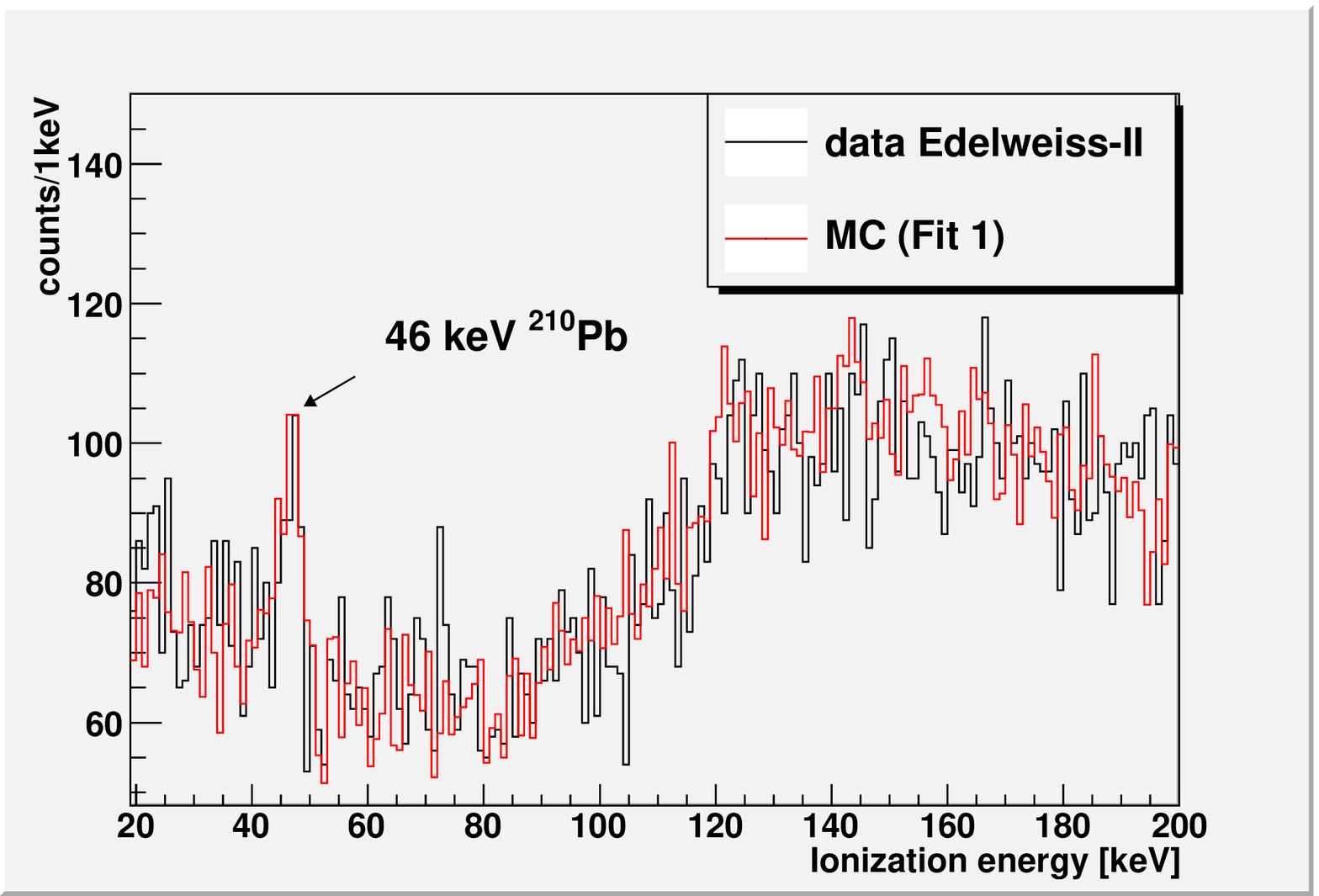}
\end{minipage} 
\caption{\label{fig:gamma_back} The background ionisation energy spectrum in the fiducial volume of the EDELWEISS-II detectors from measured data (black line) and Monte Carlo simulation (red line) for 185 kg$\times$days. The full energy range of 0--3000 keV is shown on the left and the relevant range for WIMP search (20--200 keV) is shown on the right.}
\end{figure}

\begin{table}[htb]
\centering
\caption{Ionisation event rate in events/kg/day in fiducial volume obtained from simulations.}
{\footnotesize
\begin{tabular}{lllllll|l}
\hline
 Material                &   \multicolumn{7}{c} {Gamma event rate (events/kg/day) at 20-200 keV}  \\
                         &  \multicolumn{6}{c} {Fit 1} &  Fit 2 \\
                              & $^{226}$Ra& $^{228}$Ra ($^{228}$Th) &  $^{60}$Co& $^{40}$K& Other&  Total (\%)& Total (\%)\\
	&	&	&	&	& radionuclides & & \\
                              \\
\hline
\\
Ge crystals                   &     0           &   0             &  0               &   0           &  $^{68}$Ge: 1.6 & 1.6 (2)   & 1.6 (2)    \\
\\
Detector casings/CuC2 copper     &       1.2    &  1         &    1             & 0             &$^{210}$Pb: 11    &  14 (17)  & 14 (18)  \\
\\                    
Disks, bars, 10 mK chamber/         &       0.2      &  1          &  5              & 0.3         &  $^{57}$Co: 0.7     &9.5 (12) &  13.5 (17)    \\
CuC1 copper                      &                &                &              &              & $^{54}$Mn: 2.3        &  &     \\
\\
Screens 7 to 11/copper               &       12        &   15          & 3       &  2             &  $^{57}$Co: 0.2   &   32.5(40)  &  17 (22)  \\   
                                   &               &                 &          &                 &      $^{54}$Mn: 0.3            &    \\
                                   \\  
Pollution 300K (see text)              &      8             & 14        &     0        &     0         &        0  &      22 (27) &  29 (37) \\
\\
Modern lead shield                  &       0             & 2.6        &     0        &     0         &        0  &      2.6 (3) &  4 (5) \\
\\
\hline
\\ 
Total MC                     &     21            &   33.6          &  9       & 2.3         &                    &       82   &  79   \\
\\
Total data                   &                  &                &               &              &                       &   82  & 82      \\
\\  
\hline    
\end{tabular}
}
\label{tab:eventrate_le}
\end{table}

Figure~\ref{fig:gamma_back} shows the gamma-background in the fiducial volume of the EDELWEISS-II detectors compared to the GEANT4 simulation results.  The data were collected with the EDELWEISS-II set-up containing 15 germanium detectors of the type described in Ref.~\cite{martineau} with a total exposure of 310 kg$\times$days. After a cut on the fiducial volume, data with an exposure of 185 kg$\times$days were compared to the simulations. No multiplicity cuts have been applied, i.e. coincident pulses between detectors were included. Multiple-hit events contribute about 30\% to the background rate in data and simulations and their rejection does not change the results presented here.
Some characteristic peaks are observed in the 0--3000 keV region: $^{60}$Co peaks at 1173 and 1332 keV, $^{40}$K at 1460 keV, 238 keV and 2614 keV from $^{228}$Th (the peaks are linked here to the sub-chain starting with the closest long-lived parent isotope rather than to the gamma-ray emitter).
On the right plot of Figure~\ref{fig:gamma_back} the 46 keV peak from $^{210}$Pb can be seen. 
The contributions to the gamma background in the low-energy region are presented in Table~\ref{tab:eventrate_le} for two fitting results, corresponding to the minimum and maximum contributions of the thermal cryostat screens. The primary source of gamma background is connected to the U/Th daughters and $^{60}$Co in copper screens 7 to 11 and 10 mK copper parts, which contribute between 39\% and 52\% to the total gamma-background. The second most important source (between 27\% and 37\%) is $^{226}$Ra and $^{228}$Ra decays in some detector parts at 300K which must be introduced to match the data. This source (marked as `Pollution 300K' in Table~\ref{tab:eventrate_le}) might be due to radioactivity in cryogenic pipes, bolometer boxes, uncontrolled impurities on the 300K screen or radon still present in the air in the gap between the cryostat and the lead shielding, in spite of the flushing of radon depleted air. The third most important gamma background source is the $^{210}$Pb surface pollution at the level of  the detector casings or on the detector's surface (17\%).

\section{Neutron background}
The Monte Carlo simulation used the GEANT4 High Precision (HP) model for neutrons with energies below 20 MeV. Elastic and inelastic scattering, capture and fission were included. 

\begin{figure}[htb]
\begin{center}
\includegraphics[width=10cm]{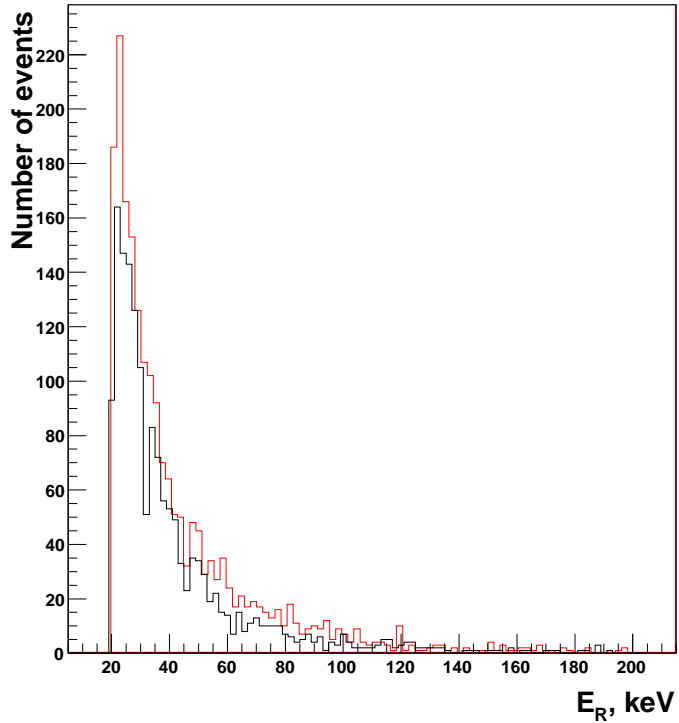}
\caption{Energy spectra of neutrons from the Am-Be source used in the detector calibration. Black histogram -- data from calibration run with the neutron source; red histogram -- spectrum obtained in the simulations of neutrons from this source. Data have not been corrected for the detector dead time of DAQ for this particular run which has been found to be $30\pm10\%$.}
\label{fig:neutrons-source}
\end{center}
\end{figure}

To check the accuracy of the model, simulations of neutrons from Am-Be source placed inside the Pb shielding on the top of the cryostat, were compared to the measured rate and energy spectrum of nuclear recoils. The source has a neutron intensity of $21\pm4$ neutrons/s and the estimated dead time of the DAQ was $30\pm10\%$. The data were collected for about 90 hours and the typical number of detected events above 20 keV after all cuts was about 2000 per crystal giving a statistical error of about 2\%. Similar statistics was accumulated in simulations. Figure~\ref{fig:neutrons-source} shows the energy spectrum of nuclear recoils observed from the neutron source and obtained in the simulations. Data have not been corrected for the dead time on this plot so lie below the simulated histogram but the overall shape of the spectra are in good agreement. The ratio of measured-to-simulated event rates above 20 keV after all cuts, corrected for the dead time and averaged over all crystals, was found to be $1.20 \pm 0.23$, where the error, given at 68\%~C.~L. is dominated by a 19\% uncertainty in the source intensity. Statistical and dead time uncertainties are also included. The ratio is consistent with 1 within errors, proving the validity of the geometrical model of the detector and neutron physics in GEANT4. The deviation of the average ratio from 1 may serve as an estimate of the uncertainty of the evaluated neutron rate if the source of background neutrons is located inside the polyethylene shielding.

Further tests of the simulation model were done with a strong neutron source giving about $2\times10^5$ neutrons/s, positioned outside the polyethylene and lead shielding. 50 cm of polyethylene should attenuate the fast neutron flux by 5-6 orders of magnitude. The neutron source was placed at several positions around the shielding to check the shielding model and neutron transport in GEANT4. The thickness of the shielding was not exactly the same for different source positions, the difference being as much as 5 cm of polyethylene. Also some small holes in the shielding are unavoidable due to pipes, readout cables, support structure etc, so special attention was paid to neutrons which could squeeze through these holes inside the shielding. To check the effect of the holes, the neutron source was also positioned close to the existing holes with pipes, cables and support beams. A difference up to a factor of 50 was observed in the data collected with different source positions and similar effect has also been found in the Monte Carlo simulations. For all source positions the rate of detected events after all cuts was found to be in agreement with the simulated rate within a factor of three with a typical uncertainty of 20\% for the measurements and simulations. Bearing in mind the challenge of building precise geometry of all shielding and detector components in GEANT4, the agreement between the measured and simulated rate within a factor of three can be considered as reasonably good. This is quite a small difference on a scale of the overall attenuation of the neutron flux by the polyethylene of 5-6 orders of magnitude (depending on the exact thickness of the shielding and neutron energy). A factor of three difference in the neutron rate (if being due to neutron attenuation in polyethylene) corresponds to a thickness of 5 cm of polyethylene. For half of the source positions tested, the difference between the measured and simulated rates does not exceed 50\%. The difference of 50\% in neutron flux attenuation by 50 cm of polyethylene was found between GEANT4 and MCNPX \cite{lemrani} showing a good agreement between the two codes on an overall scale of $10^6$ for the neutron flux attenuation.

\begin{figure}[htb]
\begin{center}
\includegraphics[width=10cm]{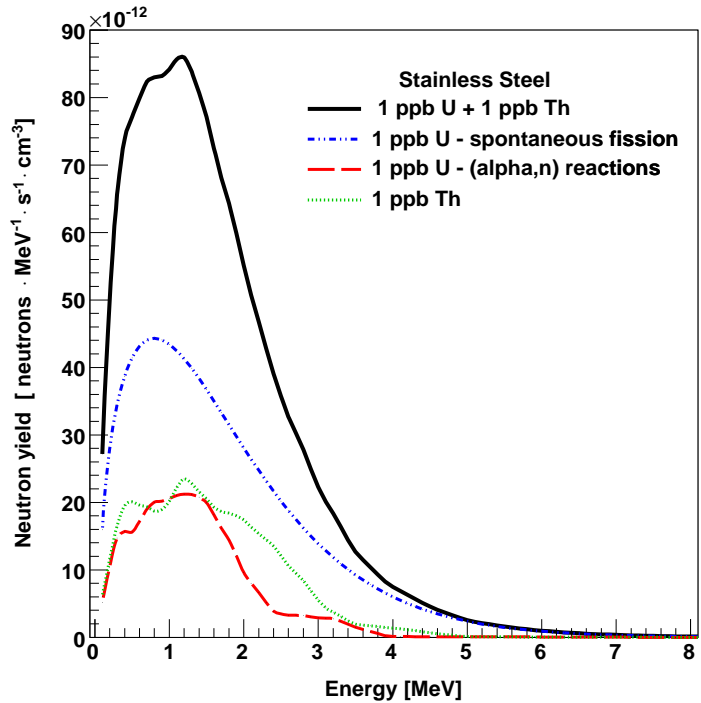}
\caption{Energy spectra of neutrons originated in U and Th decay chains in stainless steel. 1 ppb of U and Th were assumed in the calculations and resulting event spectra were later normalised to the measured concentrations. Contributions from different channels ($^{238}$U spontaneous fission and ($\alpha$,n) reactions) from the two decay chains are shown separately, together with the total spectrum.}
\label{fig:sources}
\end{center}
\end{figure}

To estimate the event rate due to neutrons in the EDELWEISS-II experiment, we considered the following materials/components as potential sources of neutrons: cavern walls (rock and concrete), lead (shielding), polyethylene (shielding), copper (cryostat and internal parts), stainless and mild steel (support structure), cables, connectors, electronic parts and other components. The results of the simulations are summarised in Table \ref{table-neutrons}. Neutron spectra were generated using SOURCES4A \cite{sources} assuming secular equilibrium in the uranium (U) and thorium (Th) decay chains. Further details about neutron production with SOURCES4A for underground experiments can be found in Ref. \cite{vito1,vito2}. Figure~\ref{fig:sources} shows the energy spectra of neutrons from U/Th decays in stainless steel generated by SOURCES4A. When calculating the neutron-induced event rate, the same cuts have been applied as to the real data: recoil energy threshold of 20 keV, ionisation energy threshold of 3 keV, 90\% acceptance in the nuclear recoil band, multiple hit and surface events have been rejected (multiple hit events have been included on the plots).

\begin{table}[htb]
\centering
\caption{Number of background events due to neutrons in EDELWEISS-II in the run detailed in ~\cite{EDW2final}. The column ``Material" refers to the material in each source which contributes most to neutron production. The column ``Composition" gives the chemical composition of the source used to calculate neutron spectra with the abundance of elements (by the number of atoms, not mass) given in brackets. Only elements with the abundance greater than 1\% are shown (with the accuracy of 1\%). The composition of the mild steel was not known so that of the stainless steel was used instead as giving slightly higher neutron flux than other possible compositions.  Neutron yield (columns 4 and 5) is shown as the number of neutrons per gram of material per second per ppb of U and Th concentration. The same cuts as for data have been applied to the simulated events. }
\vspace{0.2cm}
{\footnotesize
\begin{tabular}{llllll}
\hline
Source                       & Material     &  Composition (abundance \%)     & Neutron yield & \hspace{-0.4cm} in n/g/s/ppb  & Neutron events \\
                                    &                     &                                                           &   U   & Th  & (384 kg$\times$days) \\
\hline
Hall walls                   & Rock          & H (17), C (8), O (53), Mg (1), &  2.88$\times$10$^{-11}$ & 7.52$\times$10$^{-12}$ & $<$0.01 \cr
                                     &                    & Al (3), Si (4), Ca (13), Fe (1)  & \cr
Hall walls                   & Concrete   & H (19), C (11), O (52), &  2.21$\times$10$^{-11}$ & 3.96$\times$10$^{-12}$ &  $<$0.1 \cr
                                     &                     & Mg (1), Si (2), Ca (15) & \cr
Shielding                    & Polyethylene  & H (67), C (33) & 2.90$\times$10$^{-11}$ & 6.25$\times$10$^{-12}$ & $<$0.01 \cr
Shielding                   & Lead           & Pb (100) & 1.35$\times$10$^{-11}$ & -- & $<$0.08 \cr
Support                     & Stainless steel &  Cr (17), Mn (0.02), Fe (69),  & 1.84$\times$10$^{-11}$ & 5.92$\times$10$^{-12}$ & $<$0.01 \cr
                                    &                            & Ni (12) & \cr
Support                     & Mild steel  &  as above & 1.84$\times$10$^{-11}$ & 5.92$\times$10$^{-12}$ & $<$0.04 \cr    
Warm electronics       &  PCB    & H (22), B (2), C (19), N (6), & 7.08$\times$10$^{-11}$ & 2.21$\times$10$^{-11}$ & 1.0$\pm$0.5   \cr  
                                      &               & O (35), Mg (1), Al (4), Si (8), &   \cr
                                      &               &  Ca (3) & \cr
1K connectors               &  Aluminium  & Al (100)   & 1.80$\times$10$^{-10}$ & 8.59$\times$10$^{-11}$ & 0.5$\pm$0.2   \cr      
Thermal screens, & Copper  & Cu (100)   & 1.38$\times$10$^{-11}$ & 9.36$\times$10$^{-13}$ &  $<$0.1 \cr
crystal supports &  & &  \cr 
Coaxial cables              & PTFE   & C (33), F (67)   & 8.40$\times$10$^{-10}$ & 3.50$\times$10$^{-10}$    &  $<$0.5 \cr 
Crystal holders             & PTFE    & C (33), F (67)   & 8.40$\times$10$^{-10}$ & 3.50$\times$10$^{-10}$    & $<$0.01 \cr
Electrodes                  & Aluminium   & Al (100)  & 1.80$\times$10$^{-10}$ & 8.59$\times$10$^{-11}$    & $<$0.01 \cr
\hline                         
Total                       &      &         &   &    &   $<$3.1     \cr 
\hline     
\end{tabular}
}
\label{table-neutrons}
\end{table}

\begin{figure}[htb]
\begin{center}
\includegraphics[width=10cm]{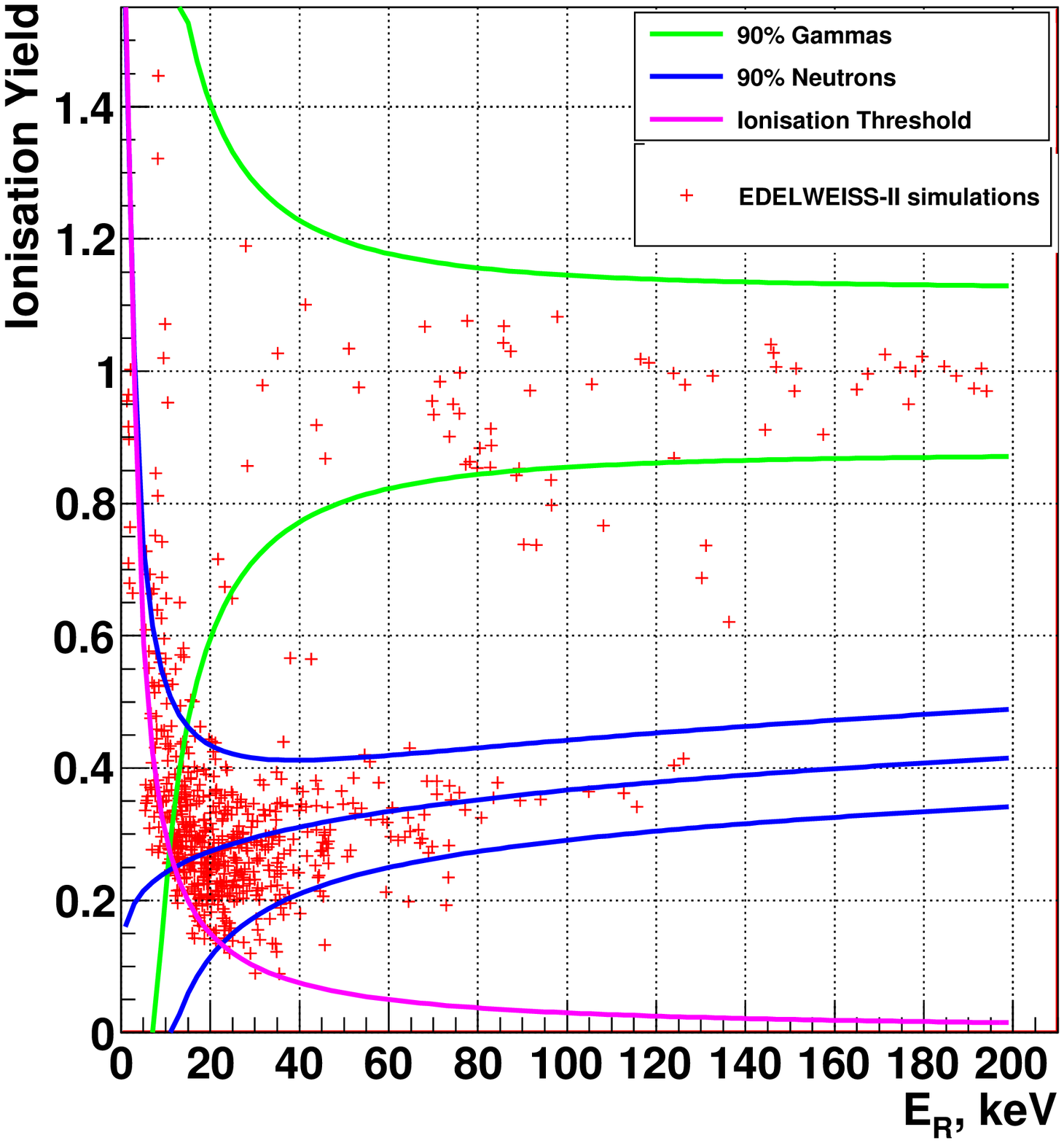}
\caption{Ionisation yield (ratio of ionisation to phonons normalised to this ratio for electron recoils) versus recoil energy for simulated nuclear recoils in EDELWEISS-II from neutrons originated in the uranium decay chain from contamination in the steel support structure around the main copper vessels. Blue curves show the average and the edges of the band which contains 90\% of nuclear recoils in one of the crystals as calculated from the experimental resolutions, that were also included in the simulations \cite{EDW2final}. Green curves show the band which contains 90\% of electron recoils. They appear on the plot because of neutron inelastic scattering and capture resulting in gamma-ray production. The pink curve shows the 3 keV software threshold for ionisation, applied as in real data. Statistics corresponds to about $4.5\times10^4$ years of live time for the uranium decay rate of 5 mBq/kg.}
\label{fig:qplot-ss}
\end{center}
\end{figure}

Apart from the mild steel, the radiopurity of different materials was taken from measurements of decay rates with Ge gamma-spectrometers or from mass-spectrometry data on U and Th concentrations. For mild steel we used the same U/Th concentrations as from the stainless steel. Note that even with 5 times higher concentrations of U/Th, the contribution of mild steel components will not exceed 0.2 events for the data reported in Ref. \cite{EDW2final}.

The measurements of the concentrations of U, Th and K (potassium) in the Modane rock and concrete (rock: $0.84\pm0.2$ ppm U and $2.45\pm0.2$ ppm Th, concrete: $1.9\pm0.2$ ppm U and $1.4\pm0.2$ ppm Th), used in the present work,  were initially reported in Ref. \cite{chazal}. 
The measurements of the neutron flux at LSM \cite{chazal} require higher values for U/Th (the normalisation requires an additional factor of 2.3 \cite{fiorucci07}) than measured in the rock/concrete due to a possible non-uniformity in U/Th abundances or rock composition. The uncertainties in the U/Th concentration, in the neutron transport through polyethylene and additional normalisation factor for the neutron flux lead to a large uncertainty (about a factor of 4.7) in the neutron event rate from the cavern walls. The upper limit on the neutron event rate from the walls is given in Table~\ref{table-neutrons} taking into account this possible error.

Since most measurements of U/Th concentrations resulted in upper limits (given at 90\%~C.~L.), normalisation of our simulation results gave upper limits on the neutron-induced event rate in EDELWEISS-II. This gives a significant contribution to the uncertainty in the neutron background event rate. The uncertainty of the neutron flux and spectra calculations using SOURCES4A has been discussed in \cite{vito1,carson} and found to be 20-30\% by comparing calculations with different cross-sections for $(\alpha,n)$ reactions and transition probabilities to excited states. The uncertainty due to the neutron transport and geometry model should not exceed 20\% (as follows from the agreement between simulations and data with a neutron source positioned within the shielding). The upper limits shown in Table ~\ref{table-neutrons} take into account all these uncertainties. The total rate shown in the last row is the sum of all upper limits and is not strictly the upper limit on the total event rate. Bearing in mind that some of the radioactivity measurements gave positive signals, we can also estimate that the lower limit on the nuclear recoil rate is 1.0 events in data reported in Ref. \cite{EDW2final}. The neutron background is potentially dominated by neutrons from materials inside the shields, especially cables and electronics.

Figure \ref{fig:qplot-ss} shows an example scatter plot of ionisation yield (ratio of ionisation energy to recoil energy normalised to this ratio for electron recoils) versus recoil energy for simulated nuclear recoils from neutrons originated in the uranium decay chain from contamination in the steel support structure around the main copper vessels. $10^6$ neutrons were sampled using the spectrum from SOURCES4A which corresponds to about $4.5\times10^4$ years of live time for the uranium decay rate of 5 mBq/kg (assuming secular equilibrium). Only events in the fiducial volume of the detectors are shown on the scatter plot. Ionisation yield, $Q$, has been calculated using the relation $Q=0.16(E_{rec}(keV))^{0.18}$, where $E_{rec}(keV)$ is the recoil energy. This relation has been proven to be valid for EDELWEISS detectors \cite{martineau,qvalue}.

To conclude, the neutron rate from radioactivity has been calculated as 1.0-3.1 events (90\%~C.~L.) at 20-200 keV in the EDELWEISS-II data run if the same cuts are applied to both data and simulations. Muon-induced neutrons  are expected to contribute $\le0.7$ events \cite{klaus}.

\section{Expected background in EDELWEISS-III}

The next stage of the EDELWEISS experiment, EDELWEISS-III is currently under construction at LSM. It will contain 40 Ge detectors (800 g each) with improved configuration of electrodes and higher fraction of fiducial mass per crystal (about 600 g) making the total fiducial mass about 24~kg~\cite{edw3}. Larger target mass requires better purity of materials close to the detectors and additional neutron shielding to reduce the expected background and achieve the projected sensitivity of a few $\times10^{-9}$ pb. Materials and components which could contribute significantly to the gamma-ray or neutron background rate in EDELWEISS-II are being replaced by their counterparts with better radiopurity, for instance the cryostat screens 7 to 11 and other copper parts at 10 mK (disks supporting the Ge detectors, vertical bars and 10 mK chamber) are made of ultra radiopure NOSV copper \cite{nosv}.

\begin{figure}[htb]
\begin{center}
\includegraphics[width=10cm]{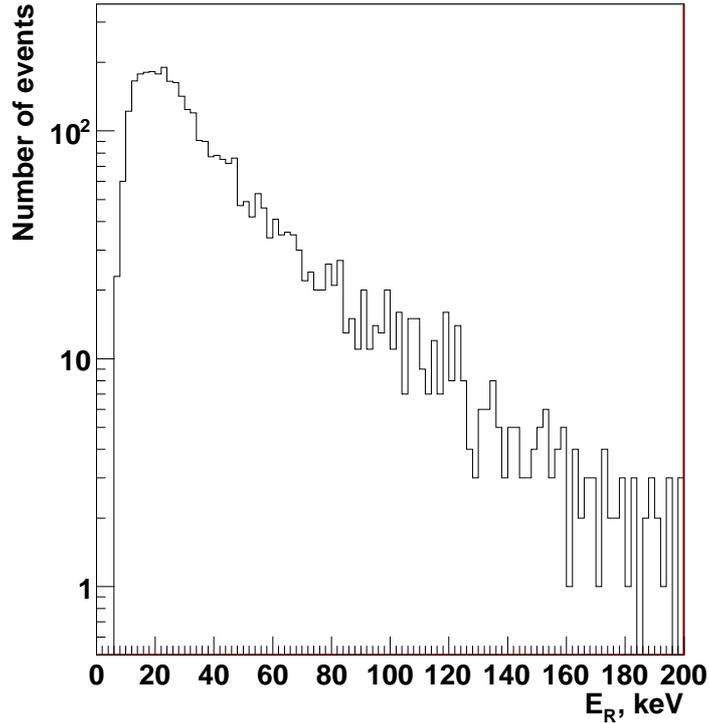}
\caption{Simulated energy spectrum of nuclear recoils in EDELWEISS-III detectors from neutrons originated in the uranium decay chain from contamination in the inner polyethylene shielding. Recoil energy has been calculated as energy deposited in a single crystal and events with all hit multiplicities have been included. Statistics corresponds to about $2.6\times10^4$ years of live time.}
\label{fig:nsp-edw3}
\end{center}
\end{figure}

\begin{figure}[htb]
\begin{center}
\includegraphics[width=10cm]{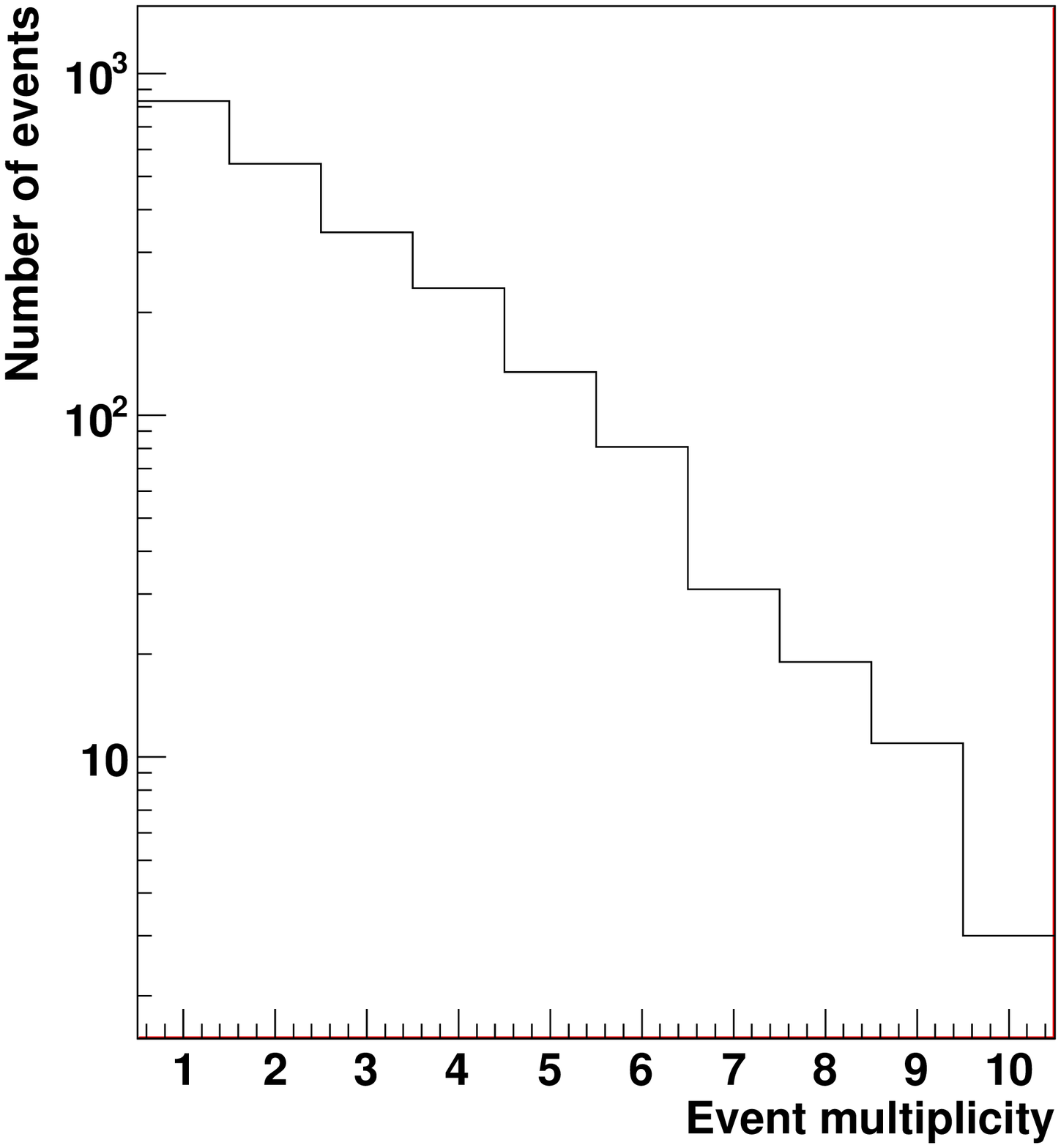}
\caption{Hit multiplicity distribution for nuclear recoils in EDELWEISS-III detectors from neutrons originated in the uranium decay chain from contamination in the inner polyethylene shielding. Statistics corresponds to about $2.6\times10^4$ years of live time. See text for details.}
\label{fig:mult-edw3}
\end{center}
\end{figure}

Radioactivity measurements of most new components were done at LSM using low-background gamma-ray spectrometry. Extensive simulations of gamma-rays and neutrons were carried out for a geometry of EDELWEISS-III with additional neutron shielding and the results were normalised to the measured concentrations of radioactive isotopes. The results of the measurements and simulations are shown in Table~\ref{table:edw3}. For some components, such as cables and connectors, various parts were screened separately using a HPGe detector. We present in Table ~\ref{table:edw3} the data for the parts which contribute the most to the background rate. The uncertainties in the radioactivity levels are given at 90\%~C.~L.. Some measurements gave only upper limits leading to large uncertainties in the expected background rates. Neutron event rates were calculated assuming secular equilibrium in the U/Th decay chains except for $^{210}$Pb sub-chain. The neutron rate is also affected by a large uncertainty in the chemical composition of the component or its part which may contribute to the background. Since a significant fraction of neutrons may come from ($\alpha$,n) reactions, exact knowledge of the chemical composition of the material is crucial in the estimate of the neutron event rate. However, in some cases, for instance electronics parts, it is not known precisely which particular part is contaminated the most and hence, it is difficult to predict the expected rate of events with high accuracy. We emphasise that we try to avoid placing materials containing elements with high cross-section of ($\alpha$,n) reactions (low energy threshold), for example fluorine, close to the crystals. As can be seen from Table~\ref{table:edw3}, a large contamination of printed circuit boards (PCBs) used in electronics (rows 8 and 9 in Table ~\ref{table:edw3})  could compromise the sensitivity of the experiment if crystals were not shielded from their radioactivity by a 14 cm thick lead plate and 10 cm of additional polyethylene shielding.

An example nuclear recoil energy spectrum for neutron events from radioactivity (uranium decay chain) in the inner polyethylene shielding and hit multiplicity distribution are shown in Figures \ref{fig:nsp-edw3} and \ref{fig:mult-edw3}, respectively. Multiplicity has been defined as the number of hits in different crystals where at least one hit was in the region of interest: recoil energy 20-200 keV, ionisation energy $>3$ keV, ionisation yield 0.1-0.5 and hit location is within the fiducial volume; other hits have only been required to have an energy higher than 10 keV. 35-40\% of events are single hit events with this selection. Energy spectrum is plotted for all events with any multiplicity. Recoil energy has been calculated as energy deposited in a single crystal. Statistics corresponds to about $2.6\times10^4$ years of live time. 

\begin{sidewaystable}
\begin{center}
\caption{\label{table:edw3}Radioactive contaminations in materials of the EDELWEISS-III set-up. All contaminations have been assessed by gamma-ray spectrometry at LSM, except for NOSV copper of thermal screens that have been taken from the measurements reported in \cite{nosv}. The last two columns give the expected gamma-induced background in events/kg/day at 20-200 keV and neutron-induced background in a year of running in 24 kg of fiducial mass. For 15 keV threshold the gamma background will change by less than 3\% whereas the neutron background will increase by about 15--20\%. The first 5 rows with data show the materials positioned close to the crystals so crystals are directly exposed to the radiation from these components. The next 3 rows show the materials below the lead plate and polyethylene beneath the detectors. A small gamma rate from warm electronics is due to the additional lead which shields the crystals from the gamma radiation. The gamma-induced rate is given for all events within the fiducial volume without excluding coincidences between different crystals. For neutron-induced rate the coincidences were excluded assuming the threshold for a second hit of 10 keV (35-40\% of events are single hit events with this selection). }
\vspace{2mm}

\begin{tabular}{@{}*{11}{l}}
\hline
Component & Material & Mass      & \multicolumn{5}{l}{Radioactivity in materials (mBq/kg) }          & Gammas & Neutrons \cr
                      &                 & (kg)        & $^{226}$Ra &  $^{228}$Th &  $^{210}$Pb& $^{40}$K & $^{60}$Co & (kg$\times$days)$^{-1} $& Events/year \cr
\hline
Cables         &  Apical, Cu & 0.2     & 26$\pm$15 & $<$50 & 346$\pm$110 & 167$\pm$126& $<$25 & 5--11 & 0.03--0.07    \cr
Connectors         &  Delrin, brass & 0.056  & 32$\pm$20 & $<$53 & 11000$\pm$1000 & 680$\pm$220 & $<$36 & 1--8 & 0.02--0.06    \cr
Screws         &  Brass & 0.1     & 4.9$\pm$1.3 & $<$3 & $<$100 & $<$40 & $<$3 & $<$1 & $<$0.003    \cr
Screens, support &  Cu & $\sim$500  & $<$0.016 & $<$0.012 & -- & $<$0.11 & $<$0.018 & $<$7 & $<$0.01    \cr 
Shielding &  CH$_{2}$ & $\sim$90  & 0.65$\pm$0.08 & 0.30$\pm$0.07 & $<$3 & $<$1 & $<$0.06 & 7-14 & 0.03--0.06    \cr \hline
Connectors &  Al, resin & 1.6 & 80$\pm$9 & 158$\pm$6 & 743$\pm$48 & 129$\pm$33 & $<$4 & 0.2--0.3 & 0.3--0.5    \cr
Cables         & PTFE & $\sim$1 & $<$35 & $<$28 & 190$\pm$40 & 440$\pm$110 & $<$19 & $<$1 &  $<$0.1     \cr
Cold electronics & PCB & 0.23 & 7800$\pm$500 & 12600$\pm$1200 & 4500$\pm$400 & 6500$\pm$1200 & $<$120 & 1--2 &  0.04--0.06     \cr \hline
Warm electronics   & PCB & - & 26500$\pm$1500$^{*}$ & 19300$\pm$1100 & 82000$\pm$5000 & 27000$\pm$3000 & - & $<$1 &  0.3--0.5     \cr \hline
Total & & & & & & & & 14--44 & 0.7--1.4 \cr \hline
\end{tabular}
\end{center}
$^{*}$ Decay rates for warm electronics are given for the whole set (not in mBq/kg).
\end{sidewaystable}

In addition to the components specified in Table~\ref{table:edw3} we expect to have less than 0.3 neutrons per year from components which were present already in EDELWEISS-II, such as lead shielding, mild and stainless steel support structure etc, bringing the total expected neutron rate to about 0.7-1.7 events per year of running. Decreasing the software energy threshold down to 15 keV will increase the expected neutron rate by 15--20\%. 

By comparing Tables \ref{tab:eventrate_le} and \ref{table:edw3} we can see that the improvement in the background event rate induced by gamma-rays as measured per unit mass and unit exposure time, will be up to a factor of 6. Even in the worst possible scenario of all contaminations being close to the 90\%~C.~L. upper limits (a factor of 2 improvement in gamma-induced event rate), we expect that better performance of the new ``fiducial inter-digitized" (FID) detectors compared to the old type ID detectors will allow us to reach the projected sensitivity of a few $\times10^{-9}$ pb to WIMP-nucleon cross-section. 

Since single nuclear recoil events from neutrons cannot be rejected by any discrimination technique, special measures have been taken in the new design to reduce possible background from neutrons, specifically, an additional polyethylene shielding will be installed in EDELWEISS-III. Our simulations (see Table~\ref{table:edw3}) show that this shielding will suppress the neutron background by more than an order of magnitude (per unit target mass) compared to the EDELWEISS-II setup. The neutron background given in Table \ref{table:edw3} corresponds to the rate per unit mass and exposure of $(0.8-1.9)\times10^{-4}$ events/kg/day in EDELWEISS-III compared to $(2.6-8.1)\times10^{-3}$ events/kg/day in EDELWEISS-II. An improvement by at least an order of magnitude will allow us to achieve the projected sensitivity with about 3000 kg$\times$days of statistics with EDELWEISS-III.

\section{Conclusions}
An extensive study of the gamma and neutron background in the EDELWEISS experiment has been performed, based on Monte Carlo simulations combined with radiopurity data. The primary source of gamma background in EDELWEISS-II is the copper from the cryostat screens and 10 mK parts. The neutron background is potentially dominated by neutrons produced by $\alpha$-n reactions in materials inside the shields, in particular cables and electronics. The calculated neutron rate from radioactivity of 1.0-3.1 events (90\%~C.~L.) at 20-200 keV in the EDELWEISS-II data run together with the expected upper limit on the misidentified gamma-ray events ($\le0.9$), surface betas ($\le0.3$) \cite{EDW2final}, and muon-induced neutrons ($\le0.7$) \cite{klaus}, do not contradict 5 observed events in nuclear recoil band \cite{EDW2final}. 
The background studies performed in the present work have contributed to the design of the next stage of the experiment, EDELWEISS-III. New cryostat screens and 10 mK parts will be built from ultra-pure copper and an inner polyethylene shielding against neutrons from materials inside the external shielding will be installed. 
The expected gamma-ray and neutron induced background rates from radioactivity in EDELWEISS-III at 20-200 keV are 14-44 events/kg/day  and 0.7-1.4 events in 40 detectors per year, respectively. With these improvements and the projected increase by an order of magnitude of the detector mass, the goal is to soon probe the range of spin-independent WIMP-nucleon cross-sections down to a few $\times 10^{-9}$ pb.

\section*{Acknowledgments}
The help of the technical staff of the Laboratoire Souterrain de Modane is gratefully acknowledged. Matthias Laubenstein has kindly measured the copper of type CuC2 using the GeMPI detector at the Laboratori Nazionale de Gran Sasso, LNGS, developed by the Max-Planck-Institut fuer Kernphysik (MPIK) in Heidelberg. The EDELWEISS project is supported in part by the Agence Nationale pour la Recherche (France) under contract ANR-10-BLAN-0422-03, the Russian Foundation for Basic Research (Russia) and the Science and Technology Facilities Council (UK, grants ST/I003371/1, ST/I00338X/1, ST/J000671/1, ST/J000663/1, ST/K003186/1, ST/K006444/1, ST/K003151/1). Background studies for dark matter search are funded in part by the German ministry of science and education (BMBF) within the "Verbundforschung Astroteilchenphysik" grant 05A11VK2.



\begin{thebibliography}{9}
\bibitem{EDW2final} E. Armengaud et al. (The EDELWEISS Collaboration). {\it Phys. Lett. B}, {\bf 702}, 329 (2011).
\bibitem{interdigit1} A. Broniatowski et al. {\it J. Low Temp. Phys.}, {\bf 151},  830 (2008).
\bibitem{interdigit2} A. Broniatowski et al. (The EDELWEISS Collaboration). {\it Physics Letters B}, {\bf 681},  305 (2009).
\bibitem{klaus} B. Schmidt et al. (The EDELWEISS Collaboration). {\it Astroparticle Physics}, {\bf 44}, 28 (2013).
\bibitem{veto} V. Yu. Kozlov et al. {\it Astroparticle Phys.}, {\bf 34}, 97 (2010).
\bibitem{chazal} V. Chazal et al. {\it Astroparticle Phys.}, {\bf 9}, 163 (1998).
\bibitem{fiorucci07} S.~Fiorucci et al. (The EDELWEISS Collaboration). {\it Astroparticle Phys.}, {\bf 28}, 143 (2007).
\bibitem{GeMPI} G. Heusser, M. Laubenstein and H. Neder. {\it Radionuclides in the Environment: International Conference on Isotopes in Environmental Studies} (Aquatic Forum 2004, 25-29 October, Monaco), ed. by P. Povinec, J. A. Sanchez-Cabeza, Radioactivity in the environment, {\bf 8} 495 (2006). 
\bibitem{XFN}  X.-F. Navick et al. {\it Nucl. Instrum. \& Meth. A}, {\bf 444}, 361 (2000).
\bibitem{Geant4} S. Agostinelli et al. {\it Nucl. Instrum. \& Meth. A}, {\bf 506}, 250 (2003).  
\bibitem{martineau} O. Martineau et al. (The EDELWEISS Collaboration). {\it Nucl. Instrum. \& Meth. A}, {\bf 530}, 426 (2004).
\bibitem{Geant4_manual} http://geant4.web.cern.ch/geant4/UserDocumentation/UsersGuides/\\PhysicsReferenceManual/fo/PhysicsReferenceManual.pdf.
\bibitem{sources} W. B. Wilson et al. SOURCES4A, Techical Report LA-13639-MS, Los Alamos (1999). 
\bibitem{lemrani} R. Lemrani et al. {\it Nucl. Instrum. \& Meth. A}, {\bf 560}, 454 (2006).
\bibitem{vito1} V. Tomasello, V. A. Kudryavtsev, M. Robinson. {\it Nucl. Instrum. \& Meth. A}, {\bf 595}, 431 (2008).
\bibitem{vito2} V. Tomasello, M. Robinson, V. A. Kudryavtsev. {\it Astroparticle Phys}, {\bf 34}, 70 (2010).
\bibitem{carson} M. J. Carson et al. {\it Astroparticle Phys}, {\bf 21}, 667 (2004).
\bibitem{qvalue} P. D. Stefano et al. {\it Astropart. Phys.}, {\bf 14}, 329 (2001).
\bibitem{edw3} E. Armengaud (for the EDELWEISS Collaboration). Talk at the 12th International Conference on Topics in Astroparticle and Underground Physics (TAUP2011, Munich, 5-9 September 2011); http://taup2011.mpp.mpg.de/
\bibitem{nosv} M. Laubenstein et al. {\it Applied Radiation and Isotopes}, {\bf 61}, 167 (2004).
\end{thebibliography}
\end{document}